\documentclass[12pt]{article}

\usepackage{amsmath}
\usepackage{amssymb}
\usepackage{amsfonts}
\usepackage{graphicx}

\numberwithin{equation}{section}

\begin{document}

\begin{center}
{\large \bf{ Evidence of the true  Higgs boson $H_T$ at  the LHC Run 2}}
\end{center}

\vspace*{2cm}

\begin{center}
{
Paolo Cea~\protect\footnote{Electronic address:
{\tt Paolo.Cea@ba.infn.it}}  \\[0.5cm]
{\em INFN - Sezione di Bari, Via Amendola 173 - 70126 Bari,
Italy} }
\end{center}

\vspace*{1.5cm}

\begin{abstract}
\noindent 
 The aim of the present note is to  compare the recent LHC data  at  $\sqrt{s} =13 \,TeV$ with
 our previous theoretical proposal   that  the  true Higgs boson $H_T$ should be a broad heavy resonance 
  with mass around  $750 \, GeV$. We focus on the  so-called golden channel  $H_T \rightarrow ZZ$ where the  pair 
  of Z bosons decay leptonically to  $\ell^+ \ell^- \ell^+ \ell^-$, $\ell$ being  either an electron or a muon. 
 We use the data collected by the ATLAS and CMS Collaborations at  $\sqrt{s} =13 \,TeV$  with
 an integrated luminosity of $36.1 \, fb^{-1}$  and $77.4 \, fb^{-1}$ respectively.
We find that  the experimental data from both the LHC Collaborations  do display in the golden channel
a rather broad resonance structure around $700 \, GeV$ with a sizeable statistical significance.
Our theoretical expectations seem to be in fair good agreement with the  experimental observations.
Combining the data from both the  ATLAS and CMS Collaborations we obtain an evidence of the heavy Higgs boson  in this channel 
with an estimated statistical significance  of more than  five  standard deviations. 
\vspace{0.5cm}

 \vspace{0.2cm}


\end{abstract}
\newpage
\noindent
\section{Introduction}
\label{s-1}
The mechanism of spontaneous symmetry breaking, now called the
Brout-Englert-Higgs  mechanism~\cite{Englert:1964,Higgs:1964,Guralnik:1964,Higgs:1966}, is a fundamental
aspect of the Standard Model Physics.
Actually, the first runs of proton-proton collisions at the CERN Large Hadron Collider (LHC) with center-of-mass
energies $\sqrt{s} = 7 \, ,\, 8 \; TeV$ (Run 1) has brought the confirmation of the existence of 
a spin-zero boson H with mass $m_H \simeq 125 \, GeV$ which resembles the one which breaks the electroweak symmetry in the Standard Model~\cite{Aad:2012,Chatrchyan:2012}.  \\
Usually the spontaneous symmetry breaking in the Standard Model is implemented
within the perturbation theory which leads to predict that the Higgs boson mass squared  is proportional to $\lambda \, v^2$,
where $\lambda$ is the renormalised scalar self-coupling and  $v \simeq 246 \; GeV$ is the known weak scale. 
 On the other hand, it is known that,  within the non-perturbative  description of spontaneous symmetry breaking in the Standard Model,
 self-interacting scalar fields are subject to the triviality problem~\cite{Fernandez:1992}, namely the renormalised self-coupling
$\lambda \rightarrow  0$ when the ultraviolet cutoff  is sent to infinity.  Strictly speaking, there are  no rigorous proof of triviality.
Nevertheless, there exist several numerical studies  which leave little doubt on the triviality conjecture.  As a consequence, 
within the perturbative approach,  the scalar sector of the Standard Model represents just an effective description  valid only up to some
cut-off scale. \\
If the renormalised self-coupling of the scalar fields vanishes, then one faces with the problem of  the spontaneous symmetry 
breaking mechanism and the related scalar Higgs boson. In fact, naively, one expects that the spontaneous symmetry breaking
mechanism cannot be implemented without the scalar self-coupling  $\lambda$. However,  in Ref.~\cite{Cea:2012},  by means 
of non-perturbative numerical simulations  of the $\lambda \Phi^4$ theory on the lattice,  it was  enlightened  the scenario where
 the Higgs boson without self-interaction  could  coexist with spontaneous symmetry breaking. 
 This means that the Higgs boson condensation triggering the spontaneous breaking of the local gauge symmetries needs to
 be dealt with non-perturbatively.  
If this is the case, from one hand there is no stability problem for the condensate ground state,  on the other hand the Higgs mass is finitely
related to the vacuum  expectation value of the quantum scalar field and it can be evaluated from first principles.
 Indeed, precise non-perturbative numerical simulations indicated that the true Higgs boson,  denoted as
$H_T$ in Ref.~\cite{Cea:2017}, is a  rather heavy resonance with mass~\cite{Cea:2012}:
\begin{equation}
\label{1.1}
 m_{H_T}  \;  =  \; 754  \; \pm \; 20 \;  \;  GeV \, .
\end{equation}
In our previous paper~\cite{Cea:2017} we elaborated  some  phenomenological consequences of the massive Higgs boson
proposal.  In particular, we discussed  the couplings of the $H_T$ Higgs boson to the massive vector bosons and to fermions,
 the expected production mechanisms, and the main decay modes.  We also attempted a quantitative comparison in the so-called golden channel
 with available  LHC data   at   $\sqrt{s} =13 \,TeV$  from  both  ATLAS and CMS Collaborations  corresponding to an
 integrated luminosity of  $36.1 \, fb^{-1}$ and $35.9 \, fb^{-1}$ respectively.  The aim of the present paper is to extend
 such a comparison to the data collected by the   CMS Collaboration  in the 2016 and 2017 runs at LHC 
 corresponding to an integrated luminosity of  $77.4 \, fb^{-1}$.    
 The main results of the present note is that  the experimental data from both the LHC Collaborations  do display in the golden channel
a rather broad resonance structure around $700 \, GeV$ with statistical significances well above four standard deviations for CMS and 
 three standard deviations for ATLAS.
Moreover, we find that our theoretical expectations seem to be in fair good agreement with the  experimental observations.
We also try a combination of  the data from  the  ATLAS and CMS Collaborations corresponding to $113.5 \, fb^{-1}$ 
 and  obtain an evidence of the heavy Higgs boson  in this channel 
with an estimated statistical significance  exceeding   five  standard deviations.  \\
 We organise the paper as follows.  In Sect.~\ref{s-2}, following  Ref.~\cite{Cea:2017}, we briefly discuss 
  the couplings of our massive Higgs boson proposal to  the Standard Model gauge fields,  the main decay channels,
 and the production mechanisms. In Sect.~\ref{s-3}  we compare our theoretical proposal with 
 the recent data collected by the ATLAS and CMS Collaborations at   $\sqrt{s} =13 \,TeV$
 in the golden channel. Finally,  our concluding remarks are relegated to Sect.~\ref{s-4}. 
\section{Physics of the $H_T$ boson}
\label{s-2}
The phenomenological signatures of the massive $H_T$ Higgs boson are determined by the couplings with 
the gauge and fermion fields of the Standard Model. 
As already argued in Ref.~\cite{Cea:2017}, the coupling of the Higgs field to the gauge vector bosons is fixed by
 the gauge symmetries. As a consequence  the couplings of
 the $H_T$ Higgs boson to the  gauge vector bosons  are the same as in perturbation theory notwithstanding the
 non-perturbative Higgs condensation driving the spontaneous breaking of the gauge symmetries. 
Given the rather large mass of the $H_T$ Higgs boson, the main decay modes are the decays into two massive
vector bosons (see, e.g., Refs.~\cite{Gunion:1990,Djouadi:2008}):
\begin{equation}
\label{2.1}
\Gamma( H_T \; \rightarrow \; W^+ \, W^-)  \; \simeq  \;  \frac{G_F \, m^3_{H_T}}{8 \pi \sqrt{2}} \;
 \sqrt{1 - \frac{4 m^2_W}{m^2_{H_T}}} \;  \bigg ( 1 - 4 \,  \frac{m^2_W}{m^2_{H_T}} + 12 \, \frac{ m^4_W}{m^4_{H_T}}
 \bigg ) \; 
\end{equation}
and
\begin{equation}
\label{2.2}
\Gamma( H_T \;  \rightarrow \;  Z^0 \, Z^0) \; \simeq \;   \frac{G_F \, m^3_{H_T}}{16 \pi \sqrt{2}} \;
 \sqrt{1 - \frac{4 m^2_Z}{m^2_{H_T}}} \; \bigg ( 1 - 4 \,  \frac{m^2_Z}{m^2_{H_T}} + 12 \, \frac{ m^4_Z}{m^4_{H_T}}
 \bigg )  \; . 
\end{equation}
 Note that for heavy Higgs the radiative corrections to the decay widths can be safely 
 neglected~\cite{Fleischer:1981,Fleischer:1983,Marciano:1988}. \\
 The couplings of the $H_T$ Higgs boson to the fermions are given by the Yukawa couplings $\lambda_f$. Unfortunately,
 there are not reliable lattice non-perturbative simulations on the continuum limit of the Yukawa couplings. If we follow the
 perturbative approximation, then the fermion Yukawa couplings turn out to be proportional to the fermion mass,
 $\lambda_f = \sqrt{2} \, m_f/v$. Thus, for heavy Higgs the only relevant fermion coupling is the top Yukawa coupling 
 $\lambda_t$. 
 On the other hand, we cannot exclude that the couplings of the physical Higgs field  to the fermions
 could be very different from  perturbation theory. Therefore, in Ref.~\cite{Cea:2017}  we introduced the parameter:
\begin{equation}
\label{2.3}
\kappa \; = \; \lambda_t^2 \; \frac{v^2}{2 \, m_t^2} \; .
\end{equation}
Obviously, in perturbation theory we have $\kappa = 1$. 
Nevertheless, in our previous  work~\cite{Cea:2017} 
we found that the experimental data pointed to  strongly suppressed  fermion Yukawa couplings. Indeed, in Ref.~\cite{Cea:2017}
we assumed  $\kappa \simeq 0$. Actually, the available experimental data are still consistent
with the somewhat weaker  bound   $\kappa  \lesssim  0.15$. \\
The width for the decays of the $H_T$ boson into a $t \bar{t}$ pairs is easily found~\cite{Gunion:1990,Djouadi:2008}:
\begin{equation}
\label{2.4}
\Gamma( H_T \rightarrow \; t \, \bar{t}) \; \simeq \; \kappa \,   \frac{3 \, G_F \, m_{H_T} m^2_t}{4 \pi \, \sqrt{2}} \;
\bigg ( 1 - 4 \,  \frac{m^2_t}{m^2_{H_T}}  \bigg )^{\frac{3}{2}}  \; . 
\end{equation}
So that, to a good approximation, the Higgs total width is given by:
\begin{equation}
\label{2.5}
\Gamma_{H_T}  \; \simeq \; \Gamma( H_T \rightarrow W^+ \, W^-)  \; + \; \Gamma( H_T \rightarrow Z^0 \, Z^0)  \; + \;
 \Gamma( H_T \rightarrow t \, \bar{t})  \; .
\end{equation}
To evaluate the Higgs event production at LHC we need the inclusive Higgs production cross section. As in perturbation
theory, for large Higgs masses the main production processes are by vector-boson fusion and gluon-gluon fusion. 
In fact, the $H_T$ Higgs production cross section by vector-boson fusion is the same as in the perturbative Standard Model calculations.
Moreover, for Higgs mass in the range $700 - 800 \; GeV$ the main production mechanism at LHC is 
expected to be  by the gluon fusion mechanism.  The gluon coupling to the Higgs boson in the Standard Model is 
mediated by triangular loops of top and bottom quarks.  Since in perturbation theory the Yukawa couplings 
of the Higgs particle to heavy quarks grows with quark mass, thus balancing the decrease of the triangle amplitude, 
the effective gluon coupling  approaches a non-zero value for large loop-quark masses. This means that for
heavy Higgs the gluon fusion inclusive cross section is almost completely  determined by the top quark.
Therefore, according to our approximations the total inclusive cross section for the production of the $H_T$ Higgs boson
can be written as:
\begin{equation}
\label{2.6}
\sigma(p \; p \;  \rightarrow \; H_T \; + \; X) \; \simeq \;    \sigma_{VV}(p \; p \;  \rightarrow \; H_T \; + \; X)
\; + \; \kappa \; \sigma_{gg}(p \; p \;  \rightarrow \; H_T \; + \; X)  \; ,
\end{equation}
where  $\sigma_{VV}$ and   $\sigma_{gg}$ are the vector-boson fusion and gluon-gluon fusion inclusive cross
sections respectively. \\
The calculations of the cross sections computed at next-to-next-to-leading 
and next-to-leading order  for    heavy Higgs boson with Standard Model-like coupling
 at $\sqrt{s} = 13 \, TeV$ can be found in Ref.~\cite{Hcross13Tev} . 
As concern the gluon-gluon fusion cross section  we found~\cite{Cea:2017} that  this cross section can be
 parametrised as:
\begin{equation}
\label{2.7}
 \sigma_{gg}(p \; p \;  \rightarrow \; H_T \; + \; X)  \; \simeq \;  
 \left\{ \begin{array}{ll}
 \;  \left (  \frac{ a_1}{ M_{H_T}} 
 \; + \; a_2 \; M_{H_T}^3  \right )  \;  \exp (-  a_3 M_{H_T})  \; \; &  M_{H_T}  \; \leq \; 300  \; GeV 
  \\
 \; \; \;  \; a_4  \;  & 300 \; GeV    \leq  M_{H_T}   \leq  400 \; GeV
  \\
 \; \;  \; \;a_4 \;  \exp \big [ - a_5 ( M_{H_T} - 400 \; GeV) \big ]  \; \; &  400  \; GeV \; \leq \; M_{H_T}
\end{array}
    \right.
\end{equation}
where $M_{H_T} $  is expressed in  GeV and
\begin{eqnarray}
\label{2.8}
a_1 \simeq 1.24 \, 10^4 \; pb \, GeV \;  \; , \; \;  a_2 \simeq 1.49 \, 10^{-6} \; pb \, GeV^{-3} \; , \;  
\nonumber \\
a_3 \simeq 7.06 \, 10^{-3} \;  GeV^{-1} \; , \;  \; a_4 \simeq 9.80 \;\, pb  \; , \hspace{2.2 cm}
\\ \nonumber
a_5 \simeq 7.63 \, 10^{-3}  \; GeV^{-1}  \; . \hspace{5.05 cm}
\end{eqnarray}
Likewise,   the  dependence of the vector-boson fusion cross section  can be parametrised as:
\begin{equation}
\label{2.9}
 \sigma_{VV}(p \; p \;  \rightarrow \; H_T \; + \; X) \; \simeq \;    \bigg ( b_1 \; + \;  \frac{ b_2}{ M_{H_T}} 
 \; + \; \frac{b_3}{ M_{H_T}^2}  \bigg )  \;  \exp (-  b_4 \;  M_{H_T} )   \; ,
\end{equation}
where~\cite{Cea:2017}:
\begin{eqnarray}
\label{2.10}
b_1 \simeq - 2.69 \, 10^{-6}  \; pb   \;  \; , \; \;  b_2 \simeq 8.08 \, 10^{2} \; pb \, GeV \; , \hspace{1.15 cm}
 \nonumber \\
b_3 \simeq - 1.98 \, 10^{4}  \; pb \,  GeV^{2}  \; \;  , \;  b_4 \simeq  2.26 \, 10^{-3} \; GeV^{-1} \; .  \; \; \,
\end{eqnarray}
To compare the invariant mass spectrum of our $H_T$ Higgs with the experimental data, we note that:
\begin{equation}
\label{2.11}
 N_{H_T} (E_{1},E_{2} )  \; \simeq \; {\cal{L}} \;  \int^{E_2}_{E_1} \; Br(E) \;  \varepsilon(E) \; \sigma(p \; p \;  
 \rightarrow \; H_T \; + \; X)  \; L_ {H_T} (E)  \;  dE  
 \;  \; , 
\end{equation}
where  $N_ {H_T}$ is the number of Higgs events in the energy interval $E_1,E_2$, corresponding to an integrated luminosity 
${\cal{L}}$, in the given channel with branching ratio $Br(E)$.   The parameter $ \varepsilon(E)$  accounts for  the efficiency 
of trigger, acceptance of the detectors, the kinematic selections, and so on.  Thus, in general
 $ \varepsilon(E)$ depends on the energy, the selected channel and the detector. 
In Eq.~(\ref{2.11})  $L_ {H_T}$ is the Lorentzian distribution:
\begin{equation}
\label{2.12}
L_ {H_T} (E) \;  \simeq  \;  \; \frac{1}{ 1.0325 \;  \pi} \; \frac{\frac{\Gamma_{H_T}(E)}{2}}{\big (E \; - \; 730 \; GeV \big )^2 \; + 
\;  \big ( \frac{\Gamma_{H_T}(E)}{2} \big )^2} \;  \;   \;  \; , 
\end{equation}
where $\Gamma_{H_T} ( E)$ is given by Eq.~(\ref{2.5}), and the normalisation is  such that:
\begin{equation}
\label{2.13}
\int^{\infty}_{0} \; L_ {H_T} (E)  \;  dE \; \; = \; \; 1 \; \; .
\end{equation}
 We  assumed a slightly smaller value for the heavy Higgs boson central mass, namely
 $m_{H_T} \simeq 730 \, GeV$ that, however, is within the  statistical uncertainties of the lattice determination Eq.~(\ref{1.1}). 
\section{Comparison with the LHC data from Run 2 }
\label{s-3}
\begin{figure}
\includegraphics[width=0.5\textwidth,clip]{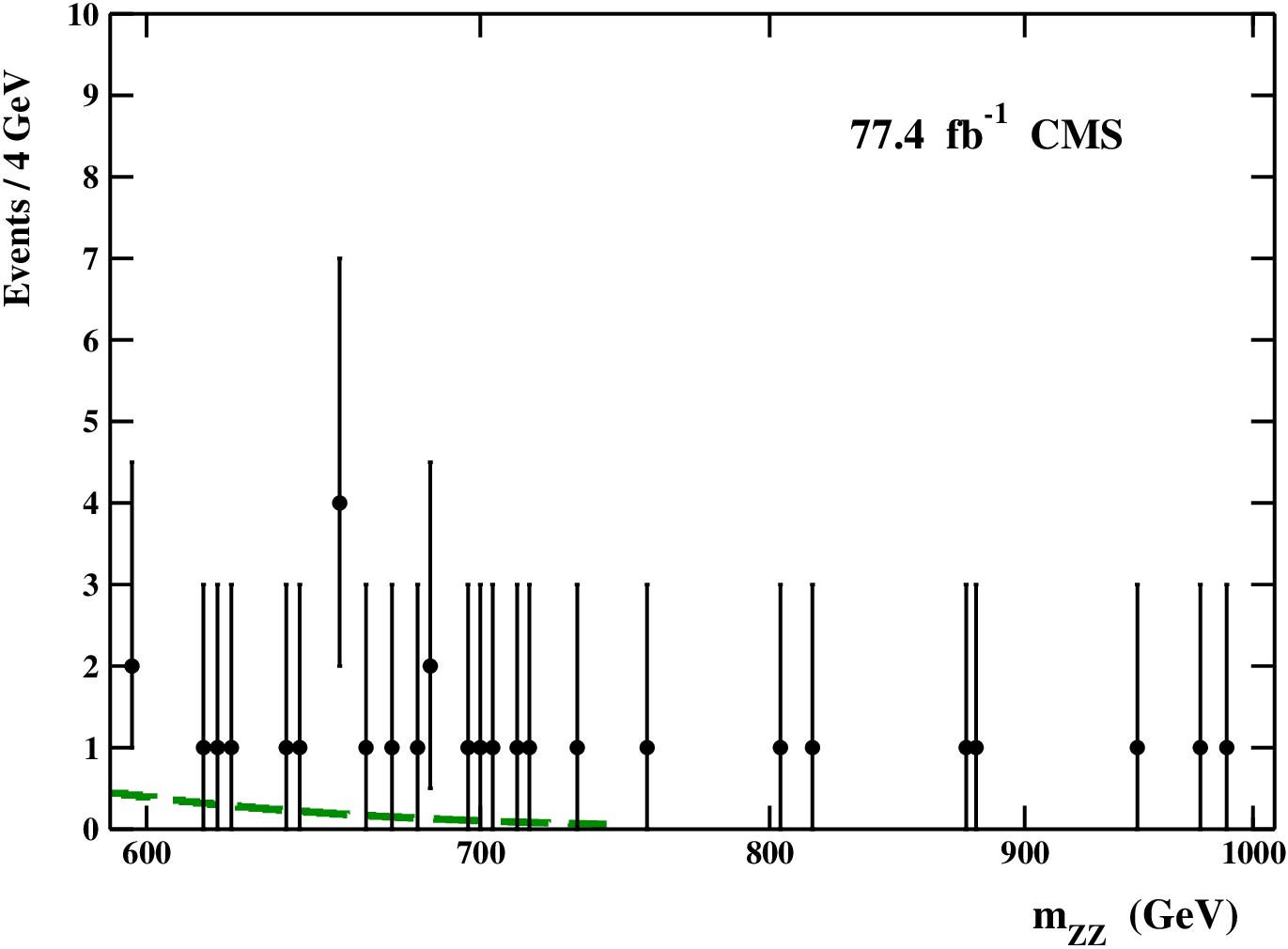}
\hspace{0.2 cm}
\includegraphics[width=0.5\textwidth,clip]{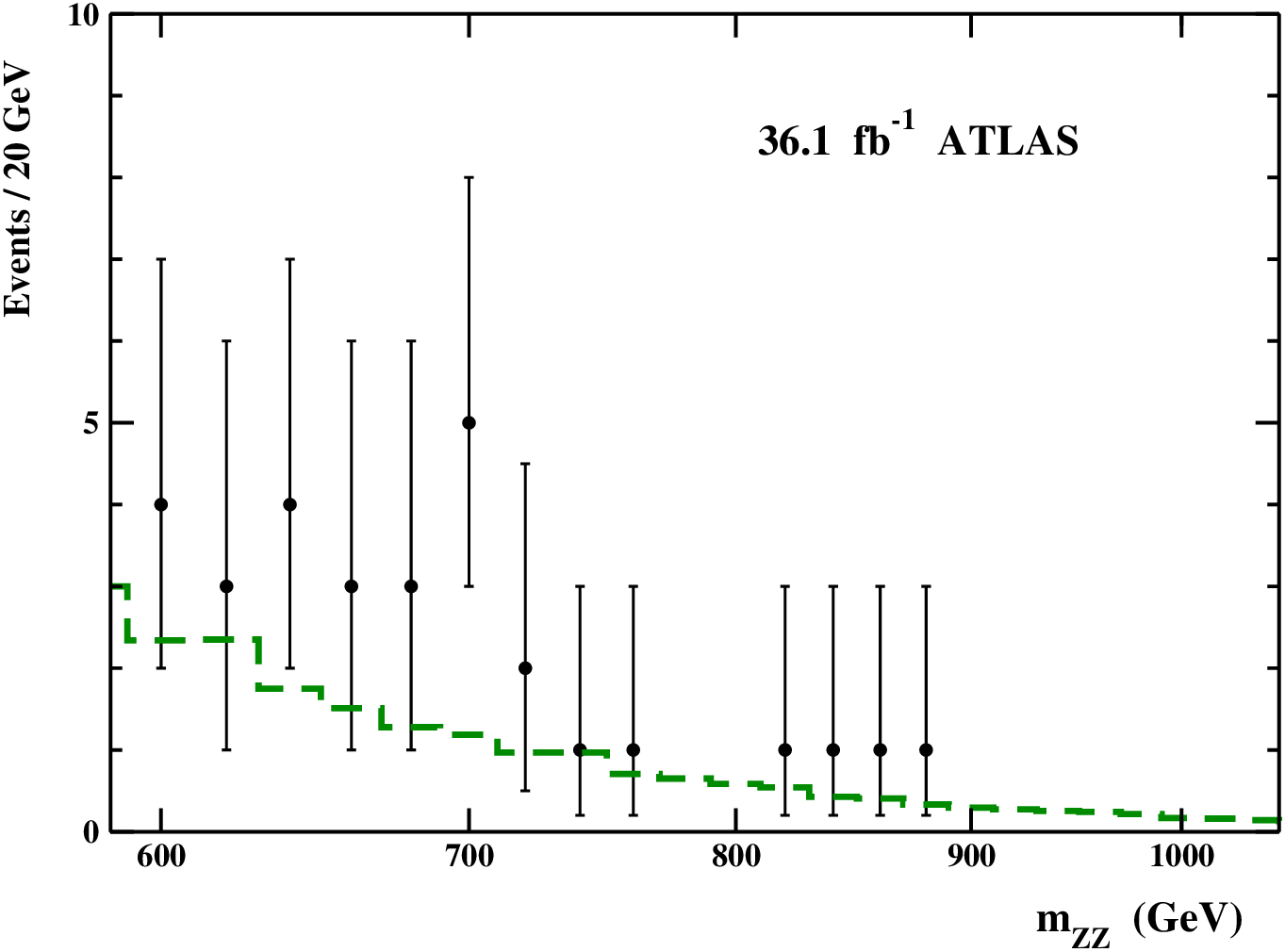}
\\
\includegraphics[width=0.5\textwidth,clip]{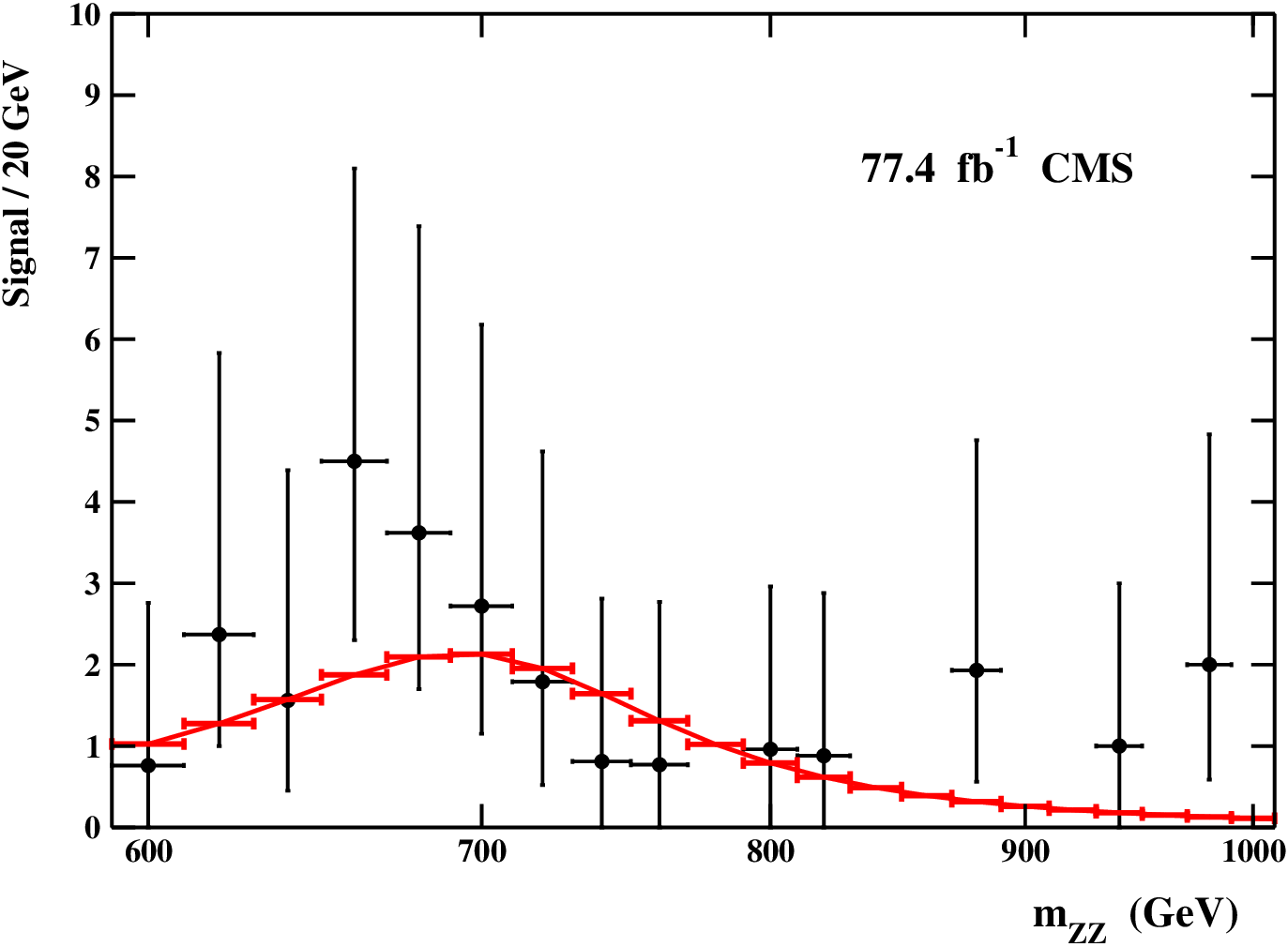}
\hspace{0.2 cm}
\includegraphics[width=0.5\textwidth,clip]{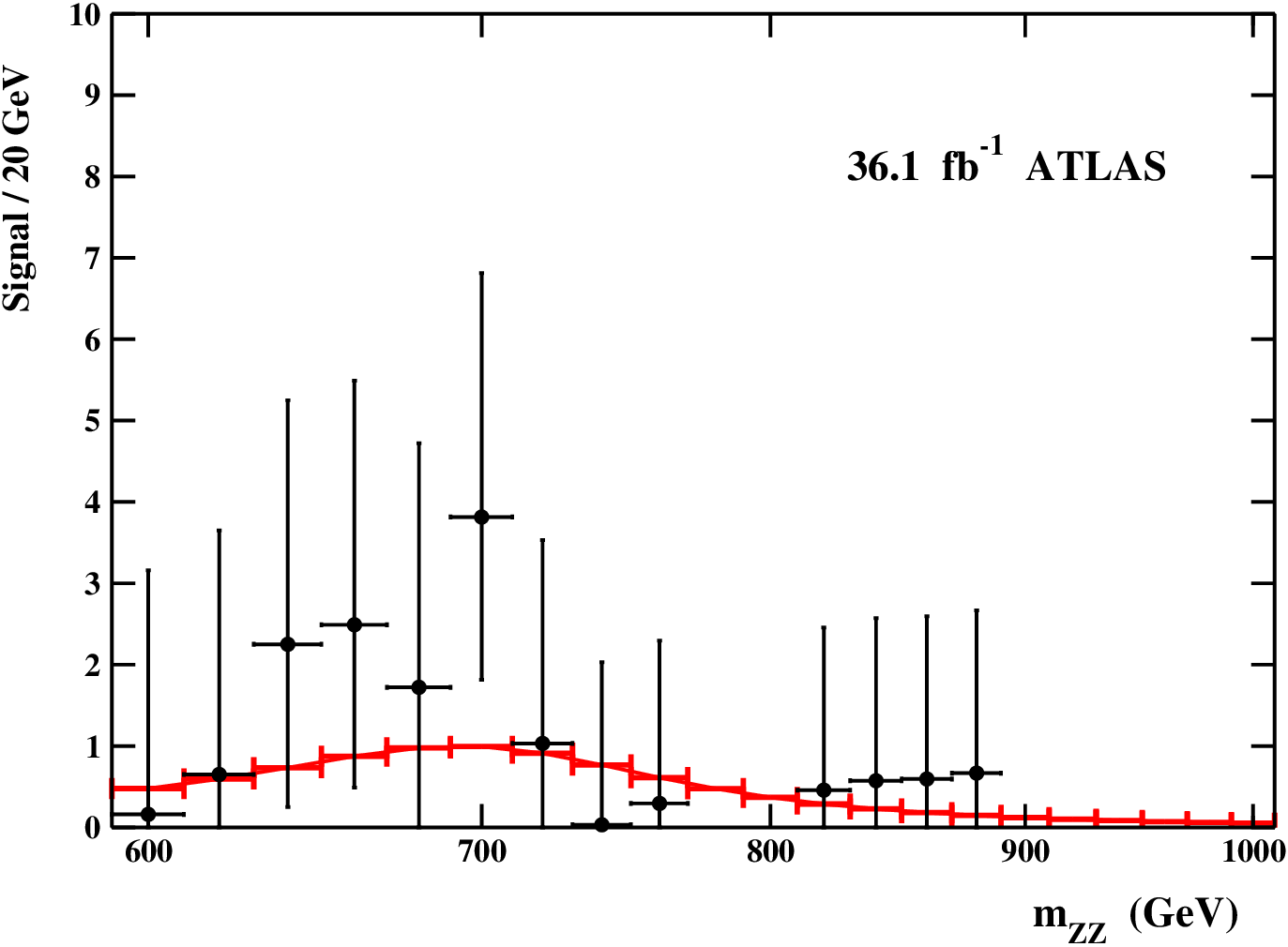}
\caption{\label{Fig1} (color online) 
Comparison to the LHC data of the distribution of the invariant mass $m_{Z Z}$ for the  process
 $H_T \; \rightarrow ZZ \; \rightarrow \ell \ell \ell \ell$   ($\ell = e, \mu$)  in the high-mass region   $m_{Z Z}  \gtrsim 600 \,  GeV$.
The CMS data (left top panel)  have been obtained  from Fig.~3, left panel, in Ref.~\cite{CMS:2017z} 
 and Fig.~2, left panel, in Ref.~\cite{CMS:2018} corresponding to a total   integrated luminosity of  ${\cal{L}} = 77.4 \,  fb^{-1}$. 
 The ATLAS data (right top panel),  corresponding  to  an  integrated luminosity of  ${\cal{L}} = 36.1 \,  fb^{-1}$,  have
 been obtained  from Fig.~4, left panel, in Ref.~\cite{ATLAS:2018}.  The dashed (green) lines are our estimate of the background.
The signal  distribution of the invariant mass $m_{Z Z}$  for CMS (left bottom panel) and ATLAS (right bottom panel).
The CMS data have been binned with bin of size  $20 \; GeV$ to facilitate  the comparison with the ATLAS data.
 The signal distributions  have been obtained from the  event distributions  by
subtracting the relevant backgrounds.  The continuum (red) lines are the  expected signal histograms obtained with
Eq.~(\ref{2.11}) by assuming  $\varepsilon(E) \simeq  0.80$ and $\kappa \simeq 0.15$.}
\end{figure}

In the present Section we compare our theoretical expectations with the available experimental data
from LHC Run 2 in the so-called golden channel corresponding to the  decays $H_T  \rightarrow ZZ  \rightarrow \ell \ell  \ell \ell $,
where $\ell$ is either an electron or a muon.
Indeed, the four-lepton channel, albeit rare, has the clearest and cleanest signature of all the possible 
Higgs boson decay modes due to the small background contamination. \\
In Fig.~\ref{Fig1} we show  the invariant mass distribution for the golden channel obtained  from the CMS experiment with
an integrated luminosity of   $77.4 \, fb^{-1}$~\cite{CMS:2017z,CMS:2018} (left top panel) and the ATLAS experiment with
an integrated luminosity of   $36.1 \, fb^{-1}$~\cite{ATLAS:2018} (right top panel).  
From our estimate of the background (dashed lines in Fig.~\ref{Fig1}) we see that, indeed, in the high invariant mass region  
 $m_{ZZ}  \gtrsim 600 \,  GeV$,  the background is strongly suppressed.  \\
 To compare  with our theoretical expectations, we display
in  Fig.~\ref{Fig1}, bottom panels,  the signal distribution of the invariant mass $m_{Z Z}$  in the golden channel. 
The signal distributions have been obtained from  the  event distributions  by subtracting the relevant backgrounds. 
 Moreover, with the aim of a comparison between the CMS and ATLAS data, we have rebinned the CMS data  with a bin of size  $20 \; GeV$. 
\\
A few comments are in order. Firstly, both the LHC experiments display almost  the same shape for the signal distributions.
Secondly, both ATLAS and CMS data do show a rather broad structure around   $m_{Z Z}  \sim 700 \,  GeV$. Finally,
the observed signal distributions seem to  compare quite well with our theoretical proposal. In fact,  in Fig.~\ref{Fig1}, bottom
panels, we display  the  expected signal histograms obtained with Eq.~(\ref{2.11}) by assuming $\kappa \simeq 0.15$ and
 $\varepsilon(E) \simeq  0.80$  to take care of the fact that the detectors do not cover the full phase space. 
To be quantitative, we may estimate  the total number of events in the invariant mass interval  
$650 \, GeV \lesssim m_{ZZ}   \lesssim  800 \,  GeV$ and compare with our theoretical expectations.
We find:
\begin{equation}
\label{3.1}
N^{obs}_{sign} = 20.74^{+9.33}_{-4.39}     \; , \; 
N^{th}_{sign} = 17.30   \; \;  \; \;   CMS \; \; \;  {\cal{L}} = 77.4 \,  fb^{-1}
\end{equation}
\begin{equation}
\label{3.2}
N^{obs}_{sign} = 13.47^{+8.73 }_{-4.14}      \; , \; 
N^{th}_{sign} =   8.10   \; \;  \;   ATLAS \;   \;  {\cal{L}} = 36.1 \,  fb^{-1}
\end{equation}
where,  to be conservative, the quoted errors have been obtained by adding in quadrature the experimental errors.
We see that CMS has an evidence of a signal reaching  a statistical significance well above four standard deviations, while
the signal significance of ATLAS is about three standard deviations. Moreover, the observed and predicted event counts
are in quite good agreement.
\begin{figure}
\centering
\includegraphics[width=0.80\textwidth,clip]{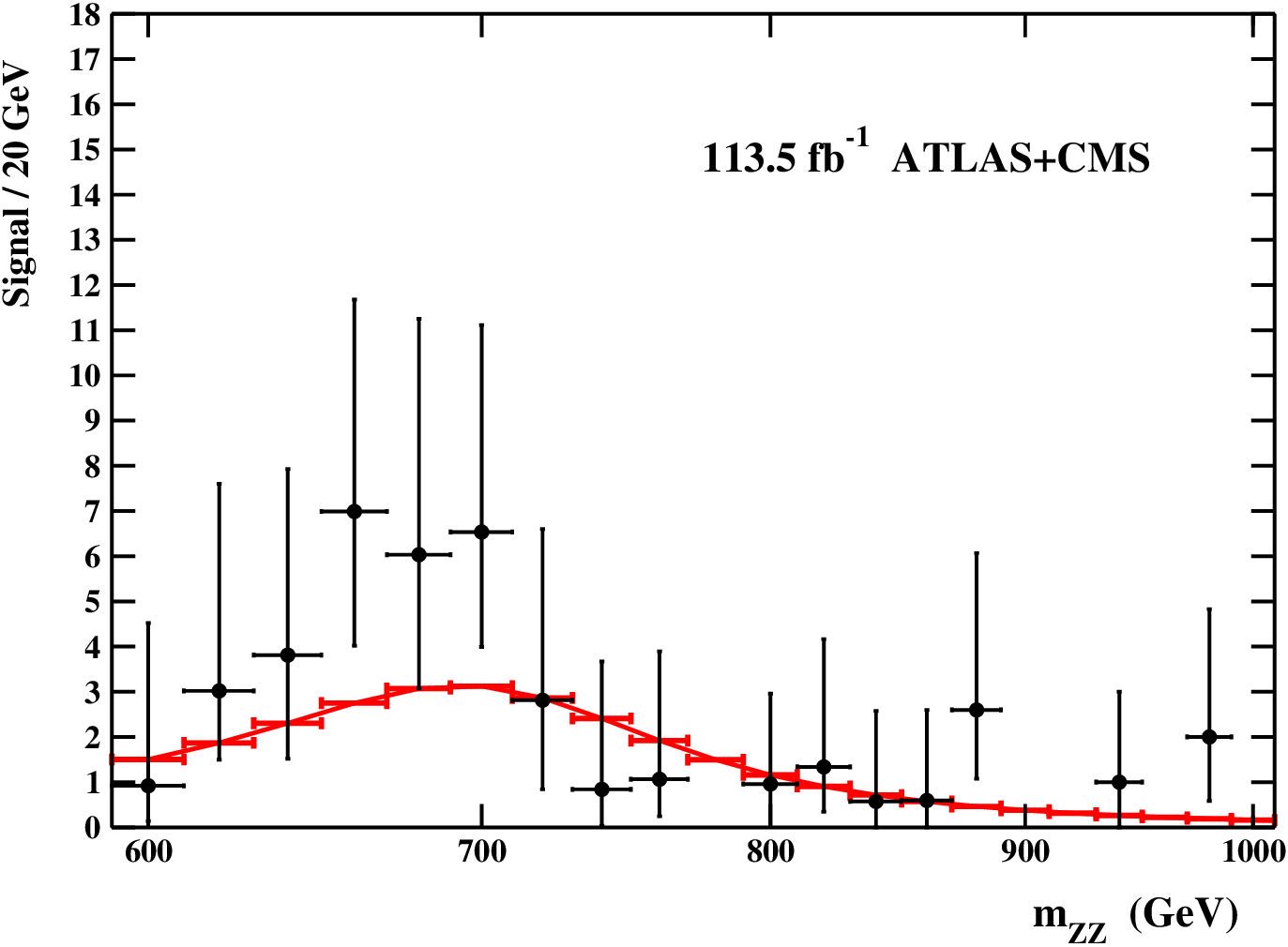}
\caption{\label{Fig2} (color online) Comparison to the LHC data of the distribution of the invariant mass $m_{Z Z}$
in the high-mass region   $m_{Z Z}  \gtrsim 600 \,  GeV$  for the  process
 $H_T \; \rightarrow ZZ \; \rightarrow \ell \ell \ell \ell$   ($\ell = e, \mu$).
The signal distribution  has been obtained from the combination of  the ATLAS and CMS event distributions  by
subtracting the relevant background.  The continuum (red) line is the  expected signal histogram assuming 
$\varepsilon(E) \simeq  0.80$ and $\kappa \simeq 0.15$.}
\end{figure}
Since the signal distributions from both the LHC experiments are compatible, we may try to combine  both LHC datasets.
In Fig.~\ref{Fig2}  we compare the resulting signal distribution with our theoretical estimates.  It is, now, evident, that
the signal distribution displays a broad peak structure around $m_{Z Z}  \sim 700 \,  GeV$ with a high statistical significance
that compare favourably  with our  theoretical signal distribution (continuum line in  Fig.~\ref{Fig2}).   
In this case we find:
\begin{equation}
\label{3.3}
N^{obs}_{sign} = 34.21^{+12.78}_{-6.03}     \; , \; 
N^{th}_{sign} = 25.40   \; \;  \; \;   ATLAS + CMS \; \; \;  {\cal{L}} = 113.5 \,  fb^{-1} \; .
\end{equation}
We see that there is an evidence of a signal reaching  a statistical significance well above five standard deviations, and
the overall observed and predicted event counts agree within $1.5$  standard deviations. 
Therefore, we may conclude that our proposal for the heavy  $H_T$ Higgs boson  is finding in the golden channel the first
clear  confirmation. 
\section{Conclusion}
\label{s-4}
It is widely believed that the new LHC resonance at $125 \, GeV$  is the Standard Model Higgs boson. 
However, stemming from the known triviality problem, i.e. vanishing self-coupling, that affects self-interacting scalar quantum fields
in four space-time dimensions, we evidenced that the Higgs boson condensation triggering the spontaneous
breaking of the local gauge symmetries needs to be dealt with non-perturbatively. It is worthwhile to notice that if this is the case,
from one hand there is no stability problem for the condensate ground state, on the other hand the Higgs mass is finitely related to the vacuum 
expectation value of the quantum scalar field and, in principle, it can be evaluated from first principles. \\
 In the present and previous papers we elaborated some phenomenological
aspects of the heavy Higgs boson scenario. We have critically discussed the couplings of the $H_T$
Higgs boson to the massive vector bosons and to fermions. We have also estimated the expected production mechanism
and the main decay modes. Comparing with the available LHC Run 2 data we concluded that the coupling of the $H_T$ Higgs
boson to fermions were strongly suppressed.
We  compared our proposal with the recent results in the golden channel from both
ATLAS and CMS Collaborations. We found that the available experimental observations were consistent with our scenario. 
We are confident that forthcoming data from LHC Run 2 will add further support to the heavy Higgs proposal. 
However, it remains the problem of unraveling the true nature of the  new LHC resonance at $125 \, GeV$. Even though we are still
convinced that the H boson cannot be the Higgs boson of the Standard Model, nevertheless up to now all the experimental informations
from the LHC experiments seem to confirm that the H boson resembles quite closely the Standard Model Higgs boson.
Thus, we see that the presence of two (perturbative and non-perturbative) Higgs bosons indicates that some fundamental and crucial aspects
of the Standard Model spontaneous symmetry breaking mechanism are still missing. 
In this respect, an interesting suggestion  has been advanced in Ref.~\cite{Castorina:2008} motivated by the stability analysis 
of the theory in a class of approximations to the effective potential that are consistent with the triviality property. Indeed, it could well
be that the elementary excitation of the scalar condensate is  a Higgs boson doublet that consists of the heavy Higgs boson 
from the upward rescaling and the a light Higgs boson  from the downward rescaling. 
 We hope that the forthcoming data from LHC will help to shed light on this fundamental aspect of the Standard Model.


\begin{thebibliography}{99}
%
\bibitem{Englert:1964}
F.~Englert and R.~Brout, Phys.\  Rev.\  Lett.\   {\bf 13},  321  (1964).
%
\bibitem{Higgs:1964}
P.~Higgs,  Phys. Lett. {\bf 12},  132 (1964).
%
\bibitem{Guralnik:1964}
G.~Guralnik, C.~Hagen and T.~Kibble, Phys.\  Rev.\  Lett.\   {\bf 13}, 585 (1964).
%
\bibitem{Higgs:1966}
P.~Higgs,  Phys. Rev. {\bf 145}, 1156 (1966). 
%
\bibitem{Aad:2012}
 The ATLAS Collaboration, G.  Aad, {\it et al.},  Phys. Lett.  B {\bf 716},   1 (2012).
 % 
\bibitem{Chatrchyan:2012}
The CMS Collaboration, S. Chatrchyan, {\it  et al.}, Phys. Lett. B {\bf 716},  30 (2012).
%
%
\bibitem{Fernandez:1992}
R.  Fernandez,  J. Fr\"ohlich,  and A. D. Sokal, 
{\em RandomWalks, Critical Phenomena, and Triviality in Quantum Field Theory}, Springer, Berlin, Germany, 1992.
%
%
\bibitem{Cea:2012}
P.~Cea and L.~Cosmai,  {\em The Higgs boson: From the lattice to LHC}, ISRN High Energy Physics, 
vol. 2012, Article ID 637950, arXiv:0911.5220.
%
%
\bibitem{Cea:2017}
P.~Cea,  {\em The $H_T$ Higgs boson at the LHC Run 2},  arXiv:1707.05605 [hep-ph].
%
\bibitem{Gunion:1990}
J.~F. Gunion, H.~E.  Haber, G. Kane, and S. Dawson, 
{\em The Higgs Hunter's Guide},  
Perseus Publishing, Cambridge, Massachusetts (1990).
%
\bibitem{Djouadi:2008}
A.~Djouadi,   Phys.\ Rept.\  {\bf 457},  1 (2008).
% 
\bibitem{Fleischer:1981}
J.  Fleischer and F. Jegerlehmer,    Phys. Rev. D {\bf 23},  2001 (1981).
% 
\bibitem{Fleischer:1983}
J.  Fleischer and F. Jegerlehmer,    Nucl. Phys. B {\bf 216},  469 (1983).
%
\bibitem{Marciano:1988}
W.~J. Marciano and S.~D. Willenbrok,    Phys. Rev. D {\bf 37},  2509 (1988).
%
\bibitem{Hcross13Tev}
https://twiki.cern.ch/twiki/bin/view/LHCPhysics/CERNYellowReportPageBSMAt13TeV.
%
%
\bibitem{CMS:2017z}
The CMS Collaboration,    {\it Measurements of properties of the Higgs boson 
 in the four-lepton final state at $\sqrt{s} = 13 \, TeV$}, 
CMS PAS HIG-16-041 (2017).
%
%
\bibitem{CMS:2018}
The CMS Collaboration,    {\it Measurements of properties of the Higgs boson in the
four-lepton final state at $\sqrt{s} = 13 \, TeV$}, 
CMS PAS HIG-18-001 (2018).
%
\bibitem{ATLAS:2018}
The ATLAS Collaboration, Eur. Phys. J. C {\bf 78}  293 (2018).
%
\bibitem{Castorina:2008}
P. Castorina, M. Consoli, and D. Zappala,   J.Phys. G {\bf 35}  075010 (2008).

%  
\end{thebibliography}
\end{document}